\documentclass[prc,twocolumn]{revtex4}
\usepackage{graphicx}
\usepackage{dcolumn}

\begin{document}
\author{D. Peterson}
\author{B. B. Back}
\author{R. V.~F.~Janssens}
\author{T. L.~Khoo}
\author{C. J.~Lister}
\author{D. Seweryniak}
\author{I. Ahmad}
\author{M. P.~Carpenter}
\author{C. N.~Davids}
\author{A. A.~Hecht}
\altaffiliation{also Department of Chemistry, University of Maryland,
College Park, MD 20742}
\author{C. L.~Jiang}
\author{T.~Lauritsen}
\author{X. Wang}
\altaffiliation{also Department of Physics, University of Notre Dame,
Notre Dame, IN 46556}
\author{S. Zhu}
\affiliation{Physics Division, Argonne National Laboratory, Argonne, IL 60439}
\author{F. G. Kondev}
\affiliation{Nuclear Engineering Division, Argonne National Laboratory, 
Argonne, IL 60439}
\author{A. Heinz}
\author{J. Qian}
\author{R. Winkler}
\affiliation{A. W. Wright Nuclear Structure Laboratory, Yale University, 
New Haven, CT 06511}
\author{P. Chowdhury}
\author{S. K. Tandel}
\author{U. S. Tandel}
\affiliation{Department of Physics, University of Massachusetts Lowell, 
Lowell, MA 01854}

\begin{abstract}
The Fragment Mass Analyzer at the ATLAS facility has been used to
unambiguously identify the mass number associated with different decay modes
of the nobelium isotopes produced via
$^{204}\mathrm{Pb}(^{48}\mathrm{Ca},xn)^{252-x}\mathrm{No}$ reactions.
Isotopically pure ($>99.7$\%) $^{204}$Pb targets were used to reduce
background from more favored reactions on heavier lead isotopes.  Two
spontaneous fission half-lives ($t_{1/2} = 3.7^{+1.1}_{-0.8}$ $\mu$s and
$43^{+22}_{-15}$ $\mu$s) were deduced from a total of 158 fission events.
Both decays originate from \thisno\ rather than from neighboring isotopes as
previously suggested. The longer activity most likely corresponds to a
$K$-isomer in this nucleus. No conclusive evidence for an $\alpha$
branch was observed, resulting in upper limits of 2.1\% for the shorter
lifetime and 3.4\% for the longer activity. 
\end{abstract}

\title{Decay modes of $^{250}$No}
\maketitle

\section{Introduction} \label{sec:introduction} 

The search for and study of the heaviest elements has yielded many surprises
and expanded our knowledge and understanding of nuclear structure. Even with
the discovery of new superheavy elements with $Z>112$, the amount of
structure information for $Z \ge 102$, though growing, remains somewhat
scarce.  It has recently been discovered that the $^{254}$No nucleus is
surprisingly robust to rotation, sustaining angular momenta up to
22$\hbar$~\cite{Ghiorso73,Reiter99,Herzberg02}. Further
studies~\cite{Herzberg02a,Hessberger04} have recently revealed the presence
of isomeric states in other No isotopes.

The reported success of using the doubly-magic nucleus $^{48}$Ca to form
superheavy elements~\cite{Oganessian99b,Oganessian00a,Oganessian04} suggests
the need to examine other reactions with this projectile.  Reactions of
$^{48}$Ca with heavier Pb isotopes have been studied in detail. Until
recently, however, there were no data available for $^{48}$Ca induced
reactions on the lightest stable Pb isotope, $^{204}$Pb. This situation was
rectified in 2001 with the systematic study by Oganessian et
al.\cite{Oganessian01} utilizing the Dubna Gas-Filled Recoil Separator
(DGFRS), and further improved in 2003 by Belozerov et al.\cite{Belozerov03}
using the VASSILISSA electrostatic separator. The DGFRS
study~\cite{Oganessian01} only observed spontaneous fission (SF) decays (no
$\alpha$ particles were seen). A half-life of $t_{1/2}=46^{+19}_{-11}\mu$s
was determined from fission events in the
$^{204}\mathrm{Pb}(^{48}\mathrm{Ca},2n)$ reaction. Events from the similar
reaction, $^{206}\mathrm{Pb}(^{48}\mathrm{Ca},4n)$, yielded a half-life of
$t_{1/2}=26^{+12}_{-6}\mu$s resulting in an average half-life of
$t_{1/2}=36^{+11}_{-6}\mu$s from 21 total decays.  The maximum observed
cross section for the $^{204}\mathrm{Pb}(^{48}\mathrm{Ca},2n)$ channel was
found to be $13^{+10}_{-7}$ nb at an excitation energy of 23.2 MeV in the
\thisno\ compound nucleus.  The VASSILISSA experiment~\cite{Belozerov03}
also observed only spontaneous fission decay. The latter measurement was
also the first to use highly-enriched $^{204}$Pb (99.6\%) material.
Forty-two events were detected with a half-life of $5.9^{+1.1}_{-0.8} \mu$s,
and were assigned to \thisno. The data also included 22 events with a
half-life of $54^{+15}_{-10} \mu$s, which were attributed to \lightno\ while
noting that the possibility of these coming from an isomeric state of
\thisno\ could not be ruled out. Neither experiment found evidence of the
250 $\mu$s SF lifetime originally assigned to \thisno\ by the authors of
Ref.~\cite{TerAkopyan75}.

The present experiment had two primary goals.  The first was to verify the
presence of multiple SF decay lifetimes in light nobelium isotopes using the
mass separation capability of the Fragment Mass Analyzer (FMA) to
unambiguously associate a specific mass with each measured decay lifetime.
The second goal was to produce a sufficient number of residues to search for
the previously unobserved $\alpha$ branch in \thisno\ in order to test
current mass models.

\section{Experimental details}
\subsection{Setup and preparation} \label{sec:setup}
The cross section for $^{204}\mathrm{Pb}(^{48}\mathrm{Ca},2n)$ at a
laboratory energy of 220 MeV is of the order of 10
nb~\cite{Oganessian01,Belozerov03}.  However, the cross sections for the
$(^{48}\mathrm{Ca},2n)$ reactions on $^{206,207,208}$Pb at 220 MeV are
30--100 times larger~\cite{Oganessian01} so that even small isotopic
impurities in the target can contribute a sizable background yield to a
measurement.  Therefore, highly-enriched ($>99.7$\%) targets of $^{204}$Pb
(0.165\% $^{206}$Pb, 0.064\% $^{207}$Pb,0.064\% $^{208}$Pb) were used.  Even
with such enrichment, residue production from these impurities could account
for up to 15\% of the total residue rate. The target consisted of four
90\degree\ Pb segments, of average thickness 540\ugcm\ sandwiched between C
entrance and exit foils of 40 and 10\ugcm, respectively,  assembled into
a wheel of 2.2 cm radius. This target wheel was rotated at 1800 rpm to
prevent deterioration of the targets due to overheating from the
high-intensity beam.

The $^{48}$Ca beam was accelerated by the ATLAS superconducting linear
accelerator at Argonne National Laboratory (ANL).  The beam intensity on
target, monitored via upstream Faraday Cups, an inductive pick-off
immediately in front of the chamber, and a Si monitor detector at
60\degrees, ranged from 40 to 90 pnA, with a time-averaged intensity of 56
pnA for the entire experiment of 3.5 days.  To further reduce the heat load
on the targets, the beam spot was slightly defocused by reducing the fields
in the last quadrupole before the target chamber.  This increased the beam
spot size from ~0.5 mm diameter to $\sim$1 mm in diameter.  The beam was
also deflected up to $\pm$2.5 mm in the vertical direction using a magnetic
steerer coupled to a triangle-wave generator.  The vertical deflection does
not affect the mass resolution since the FMA dispersion is in the horizontal
plane.  (The modest defocusing does affect the final resolution, however.)
Finally, to avoid unnecessary background due to beam scattering from the
four target wheel spokes, the beam was chopped in synchronization with the
wheel position.

Since the charge state distributions of evaporation residues (EVRs) produced
in heavy ion reactions are quite broad, the FMA was set to a fractional
charge state to allow the simultaneous collection of two neighboring charge
states (either $q$=19-20 or $q$=20-21).  The mass-to-charge ratio of the
EVRs was measured at the focal plane of the FMA with a parallel grid
avalanche counter (PGAC) filled with isobutane. The transmitted ions then
implanted into a 4 cm square, 40x40 double-sided silicon strip detector
(DSSD) positioned 40 cm downstream of the PGAC.  The data acquisition
readout was triggered on any DSSD signal above threshold, typically
150--200 mV ($\sim$300 keV).

Each DSSD signal was routed to three separate amplifiers. One signal was
passed through a delay-line amplifier to eliminate the energy ``pile-up''
characteristic of fast decays (implant to decay time of a few $\mu$s). These
amplifiers saturated (full-scale range) at 20 MeV $\alpha$-equivalent
energy.  A second branch was sent through a shaping amplifier, also with a
full-scale range of 20 MeV.  These two signals are useful for residues
implanted into the Si detector and for $\alpha$-like decay energies.  The
final branch was sent to another shaping amplifier with a full scale of
$\sim$250 MeV which recorded fission-like energies. The minimum observable
correlation time, which is related to the response of the delay-line
amplifiers, was found to be 2 $\mu$s for $\alpha$-like decays. Fission-like
energies saturated the delay-line amplifiers and only implant-fission
correlations longer than 5 $\mu$s could be observed. All events were tagged
with the readout from a 100 ns clock to provide the time differences
necessary for reconstructing correlations.  

Although the DSSD had 40 strips on each side, three strips on each side were
malfunctioning.  In addition, a fourth strip on the front side was missing
from the implant ADCs because of a faulty module.  This reduced the number
of pixels available for the detection of implant-decay correlations from
1600 to 1332.

\subsection{Calibration} \label{sec:calibration}
The high-gain channels (0--20 MeV) of the DSSD strips were calibrated with a
$^{240}$Pu-$^{244}$Cm source.  For these $\alpha$ particles, the energy
resolution was typically 15 keV FWHM, and no worse than 40 keV.  The
low-gain channels (0-250 MeV) were calibrated relative to the high-gain ones
by use of a precision pulser to generate signals over a range equivalent to
5--500 MeV.

To calibrate the FMA and ensure that the electronics were functioning as
planned, a test reaction of $^{174}\mathrm{Yb}(^{48}$Ca,$xn)$ was used to
produce $^{219,220}$Th residues. $^{219}$Th has a 1 $\mu$s half-life, making
it a suitable benchmark for fast decay events. $^{220}$Th has a 10 $\mu$s
half-life. Data from this setup reaction were collected for an approximate
beam dose of $10^{15}$ particles over nine hours.

\section{Results} \label{sec:results}
\subsection{The Ca+Yb setup reaction} \label{sec:resultsCaYb}
\begin{figure}
\includegraphics[width=.8\columnwidth]{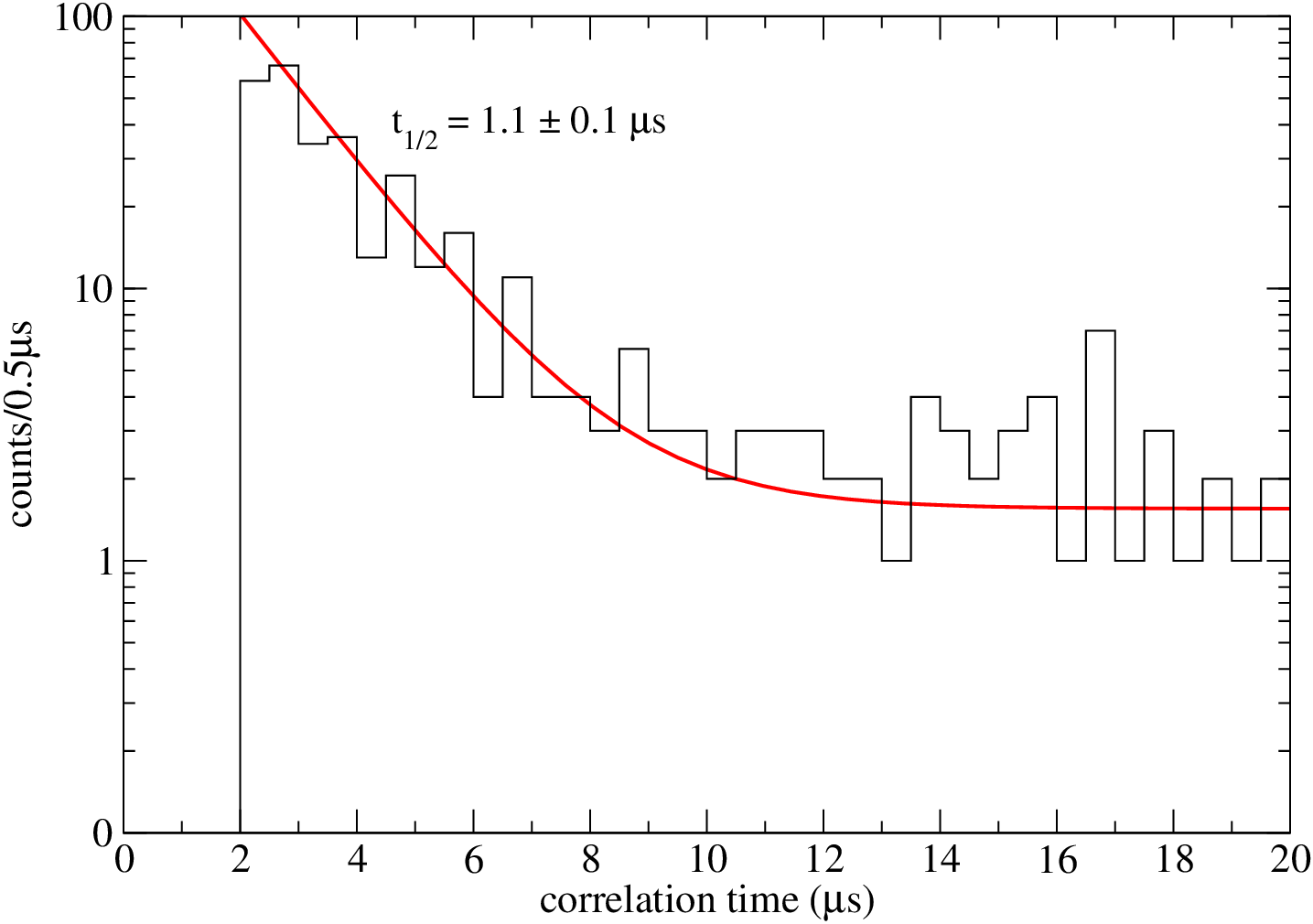}
\caption{Decay spectrum (histogram) for $^{219}$Th created in the setup
reaction and the associated fit (solid line).\label{fig:thdecay}}
\end{figure}
The setup reaction of 220 MeV $^{48}$Ca + $^{176}$Yb mainly yields Th
isotopes of mass 219 and 220.  With the FMA set for $m_0=220$ and
$q_0=19.5$, several EVRs were collected and their subsequent $\alpha$ decay
was observed.  The decay curve measured for $^{219}$Th is presented in
Figure~\ref{fig:thdecay}. Half-lives of $t_{1/2}=1.1$ and $t_{1/2}=10.5$
$\mu$s with respective uncertainties of about 10\% and 36\% were measured.
These are in good agreement with the known~\cite{Audi03} values of
1.05$\pm$0.03 and 9.7$\pm$0.6 $\mu$s, respectively, for the two isotopes.
These results confirm the proper functioning of the FMA setup. The decay
curve of \fref{fig:thdecay} also illustrates the effective minimum
correlation time for reliable recoil-alpha identification of 2 $\mu$s.

\subsection{The Ca+Pb reaction} \label{sec:resultsCaPb}
\subsubsection{Experimental details} \label{sec:details}
A total beam dose of $6.4\times10^{18}$ $^{48}$Ca particles was delivered to
the targets over the course of the experiment. Data were collected with two
different charge state settings for the FMA, $q_0=$ 19.5 and 20.5, to
collect EVRs with charge states $q=$ 19--21. The bulk of the data (98\%)
were taken at the $q_0=20.5$ setting since the focal plane yield/pnA was
highest for this setting. The average rate (scattered beam and residues)
incident on the entire DSSD was 5 events per second per pnA of incident
beam.

Two different time windows were utilized to search for implant-decay
correlations: a 500 $\mu$s window sensitive to the expected short decay time
of the primary nobelium residues, and a longer 10 s window that emphasized
both the heavier No residues produced from target impurities as well as
their fermium daughters, which also have longer lifetimes.  The analysis of
observed decays from these impurities and random correlations are discussed
in the following sections. An implant was defined as a signal in the DSSD
associated with a time-of-flight (TOF) signal between the focal-plane PGAC
and the DSSD, and located within a two-dimensional gate for residues in a
plot of $E_\mathrm{DSSD}$ vs. TOF. DSSD signals in which the high-gain
shaping amplifiers or delay-line amplifiers were saturated and the
calibrated energy in the low-gain amplifiers was greater than 50 MeV were
interpreted as originating from fission.  An energy threshold for fission
was necessary since, for the very short decay times, residue implants and
their decays occur in the same event and some implanting residues saturated
the high-gain amplifiers. These could be discriminated from fission, though,
as the residues only registered a signal of 20--40 MeV, whereas fission
events were typically above 50 MeV.  Signals from the front of the DSSD with
an energy between 6.5 and 11.5 MeV were taken to be associated with $\alpha$
particles. Table \ref{tab:eventrates} lists the observed rates for each of
these types of signals for various periods of the experiment.  The
difference in $\alpha$ rates before and after run 16 is mainly due to the
presence of the 14.6 hr $^{211}$Rn, a decay product of the $^{219}$Th
produced in the setup reaction. Runs 6--16 immediately followed the setup
reaction and a higher-than-desired background was observed as a result.
After a 48 hour period of down time, which allowed this activity to decay
away, the experiment resumed with runs numbered 18 and above. The implant
rate is directly proportional to the average beam current on target, which
steadily improved over the course of the experiment. In total, 158
recoil-fission correlations were observed.
\begin{table}
\caption{Event rates (events/sec/pixel) for each type of DSSD
signal examined---implants, fissions, and $\alpha$-like decays observed in
shaping amplifiers (SA) and delay-line amplifiers (DA). The rightmost
columns list the total number of expected random correlations (residue and
decay occurring within $500$ $\mu$s) per pixel for each run segment.
\label{tab:eventrates}}
\begin{ruledtabular}
\begin{tabular}{c|rrrr|rr}
Run & Implant & Fission & $\alpha$ (SA) & $\alpha$ (DA) &
\multicolumn{2}{c}{Expected Randoms} \\
Set & $\times 10^{-6}$ &$\times 10^{-6}$ &$\times 10^{-6}$ &$\times 10^{-6}$ & $\mathrm{EVR-F}$ & $\mathrm{EVR-\alpha}$\\ \hline
06-11   & 22.2 & 3.3 & 51.1 & 50.5 & $1.5\times 10^{-9}$ & $2.4\times 10^{-8}$ \\
12-16   & 15.4 & 4.3 & 35.8 & 34.6 & $1.0\times 10^{-9}$ & $8.0\times 10^{-9}$ \\
18-21   & 39.8 & 10.3 & 5.8 & 5.6 & $3.8\times 10^{-9}$ & $2.1\times 10^{-9}$ \\
22-37   & 97.2 & 12.2 & 2.0 & 1.9 & $1.0\times 10^{-7}$ & $1.8\times 10^{-8}$ \\
38-43   & 162.4 & 20.8 & 1.1 & 1.1 & $1.7\times 10^{-7}$ & $9.0\times 10^{-9}$ \\
\end{tabular}
\end{ruledtabular}
\end{table}

\begin{figure}
\includegraphics[width=.75\columnwidth]{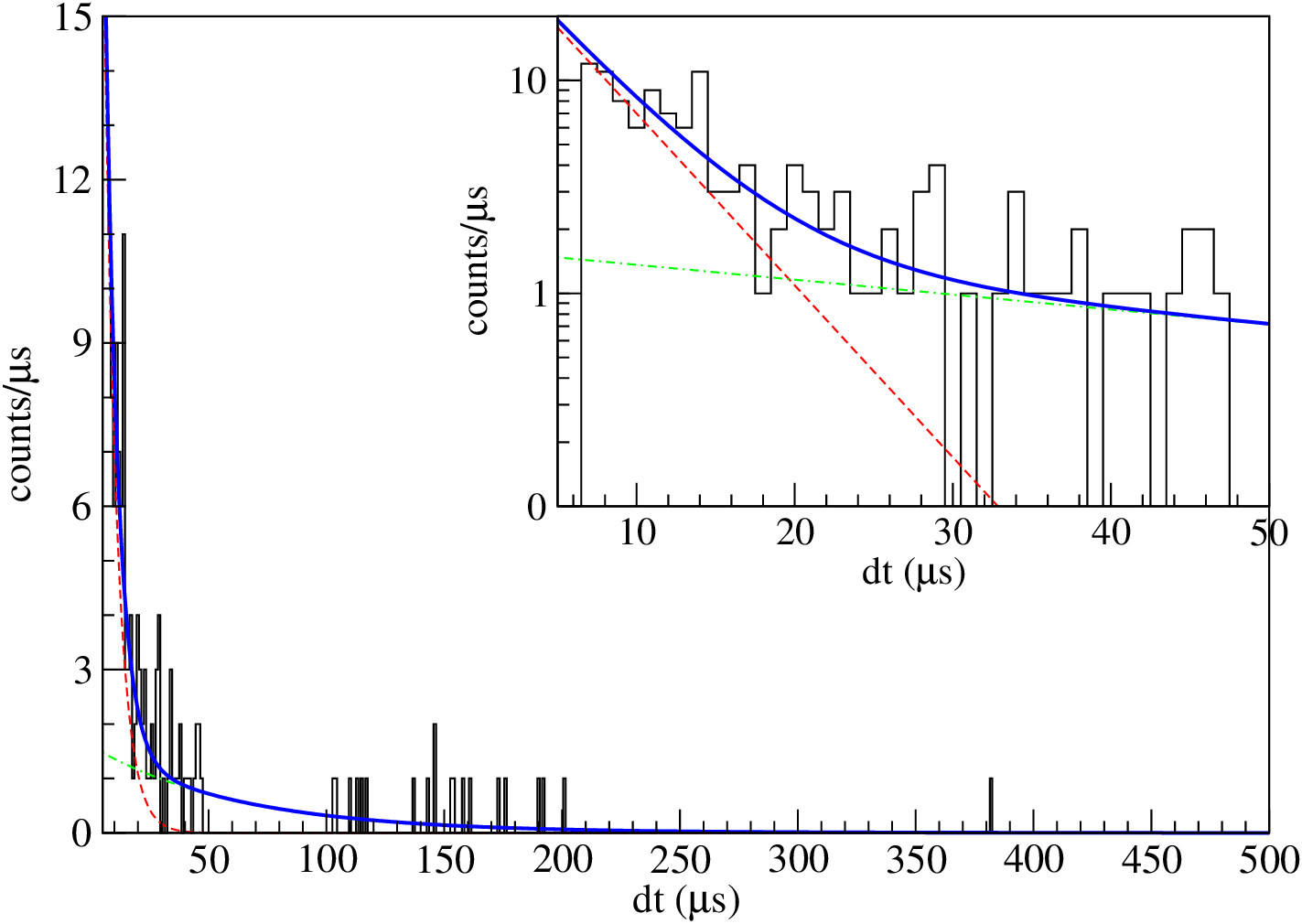}
\caption{Decay spectrum for all recoil-fission correlations observed in this
experiment. The inset expands the first 50 $\mu$s.  The solid line is the
result of a two-component fit as discussed in the text.  The dashed and
dot-dashed lines represent the contributions of the individual decay
components of half-lives 3.7 $\mu$s and 43 $\mu$s. Note that the mean flight
time through the FMA of 1.8 $\mu$s was added to the implant-decay
correlation time before binning the data.\label{fig:SFdecays}}
\end{figure}
The observed decay spectrum, given in \fref{fig:SFdecays}, illustrates
several important features. Bins of 1 $\mu$s were chosen for display and
analysis.  The full range of 500 $\mu$s matches the correlation window used
in the analysis.  The spectrum is remarkably clean and two decay components
are clearly visible. The absence of data between 45 and 95 $\mu$s is due to
the deadtime during acquisition readout.  The inset expands the first 50
$\mu$s to more clearly show the components.  The dashed, dot-dashed, and
full lines represent the results of a two-component fit described in the
following section.

\subsubsection{Fission decay details} \label{sec:fissdecay}
The key issue is whether the two fission decay components are from different
isotopes or different levels of the same No residue. In the first step of
the analysis, a crude cut between the two components was made. Decays $< 7
\mu$s are dominated by the short component, while decays occurring after 95
$\mu$s are solely due to the long component.

\begin{figure}
\includegraphics[width=.8\columnwidth]{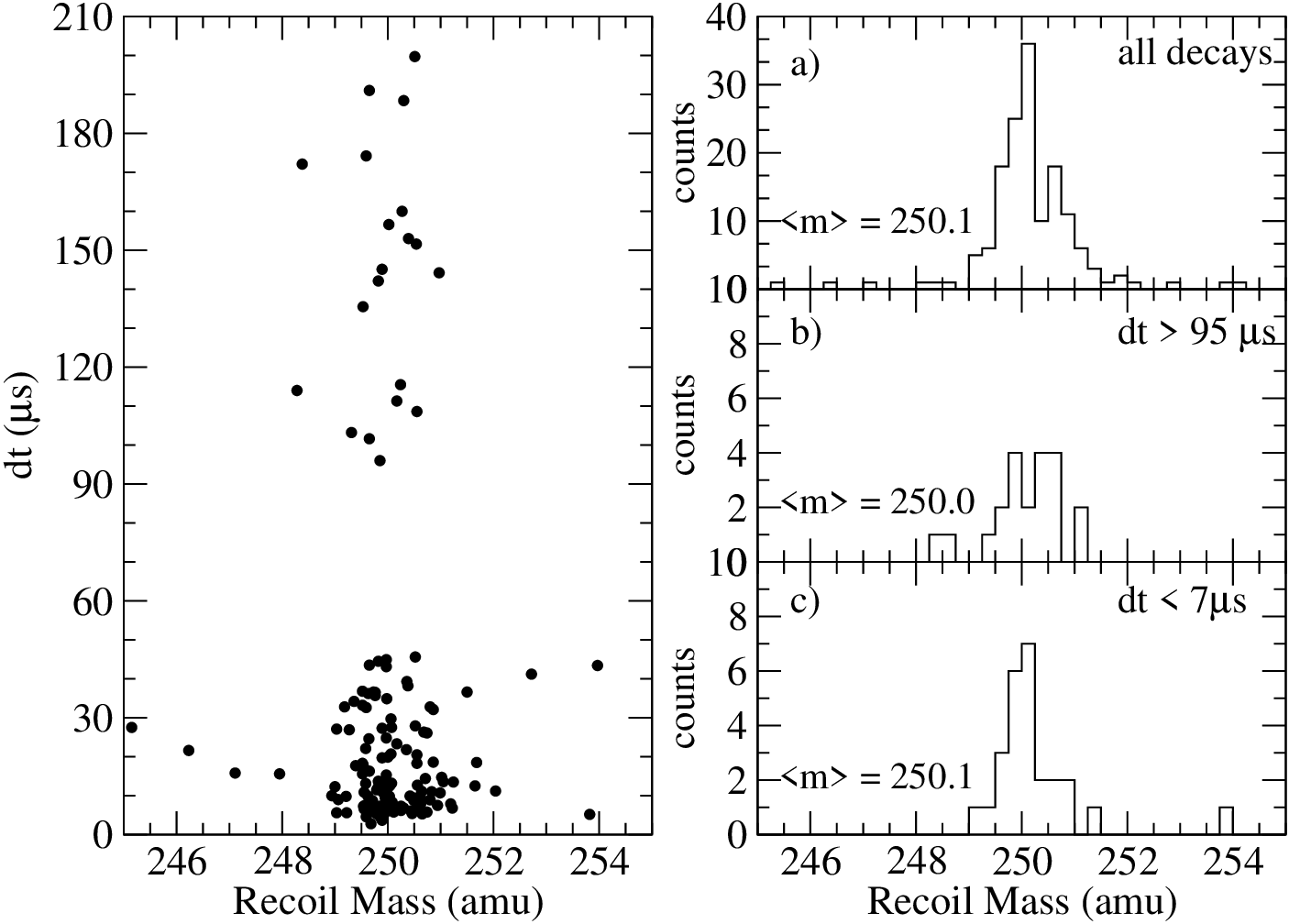}
\caption{Mass numbers of the residues associated with spontaneous fission
decays.  Left panel: decay time versus mass. Right panels: projections to
the mass axis for all decays (a), purely long decays (b), and primarily
short decays (c). See text and \tref{tab:masscent} for more
details.\label{fig:cleancut}}
\end{figure}

The masses are obtained from the focal plane (PGAC) position using the known
$m/q$ dispersion of the FMA.  The derived masses are presented in
\fref{fig:cleancut}. The left panel plots the implant-decay correlation time
versus the mass number of the implanting residue.  The right panels contain
the projections on the mass axis for all decays (a), decays longer than 95
$\mu$s (b), and decays shorter than 7 $\mu$s (c).  It should be noted that
even though there is a small contribution from the long component in the $<7
\mu$s projection, the latter is primarily dominated by the short component,
as can be seen clearly in the inset of \fref{fig:SFdecays}.  From the
distributions in \fref{fig:cleancut} it is clear that, regardless of decay
window, the distributions are similar in shape and centroid. The mean
centroid values, listed in \tref{tab:masscent}, indicate that all
decays originate from the same mass number.
\begin{table}
\caption{Centroids of the mass distribution obtained from different cuts in
the decay spectrum.  See text for details.\label{tab:masscent}}
\begin{ruledtabular}
\begin{tabular}{rrr}
Decay type & counts & centroid (amu) \\ \hline
All & 158 & 250.1 \\
$> 95 \mu$s & 20 & 250.0 \\
$< 7 \mu$s  & 31 & 250.1 \\
\end{tabular}
\end{ruledtabular}
\end{table}
Admittedly, \fref{fig:cleancut} only reveals that all decays originate from
\emph{mass number} 250 and does not provide a $Z$ identification.  However,
at low excitation energy, the $pn$ channel leading to $^{250}$Md is much
smaller in cross section than the $2n$ channel leading to \thisno.
Furthermore, the decay of $^{250}$Md is well-known\cite{Tuli05} and does not
contain a SF component. Thus, it is concluded that the two decay modes
reported here belong to $^{250}$No and not to the deexcitation of two
different isotopes, as originally suggested by the authors of
Ref.~\cite{Belozerov03}, although the possibility of these decays
originating from different states of the same nucleus was not completely
ruled out in that work.  The nature of these decays will be explored further
in section~\ref{sec:discussion}.

Extracting the lifetimes of the two decay modes is complicated by the
deadtime gap between 45 and 95 $\mu$s and by understanding the
``background'' that the long component provides to the short one. Therefore,
the long component was examined first following the procedure outlined in
Ref.~\cite{Schmidt84}. Any decays occurring after the deadtime window are
due solely to the long decay component. For a single decay component
governed by Poisson statistics, with a solid deadtime window beginning at
time-zero of width $\Delta t$, the mean lifetime, $\tau$, is the mean of the
observed decay times, $\langle dt \rangle$, minus the deadtime: $\tau =
\langle dt \rangle - \Delta t$. For the 20 decays longer than 95 ~$\mu$s,
this results in a mean lifetime of $\tau=62^{+18}_{-11} \mu$s. To estimate
the contribution of the long decay mode in the region where the short
component dominates, the amplitude, $A_\ell$ is found from the relation \[ 
\int_{95}^\infty A_\ell e^{-t/62} \dif t = 20.  \] Simply subtracting this
contribution from the decay spectrum before fitting over the first 40 $\mu$s
yields a mean lifetime of $\tau = 6.9$ $\mu$s for the short component.
However, the 20\% uncertainty in the long component decay constant results
in a factor of 2 uncertainty in the amplitude of the short component and it
complicates the determination of the correct value for its decay constant.
To better understand the uncertainties, a least-squares fit to the data was
performed with four parameters corresponding to the amplitude and decay
constant of the two components. Asymmetric errors were then computed using
the Minos routine as implemented in the code MINUIT~\cite{James75}. The
fitted mean lifetimes of $\tau=5.4^{+1.6}_{-1.2}$ $\mu$s and
$\tau=62^{+32}_{-22}$ $\mu$s (corresponding to half-lives of
$t_{1/2}=3.7^{+1.1}_{-0.8}$ $\mu$s and $t_{1/2}=43^{+22}_{-15}$ $\mu$s) are
reasonably close to those obtained from the simple estimates. The long
lifetime is the same, as should be expected since the region examined only
contains this component. The short lifetime changes somewhat, as may be
expected from the uncertainty associated with the underlying ``background''
from the long component. The resulting errors are slightly larger than those
obtained from the simple estimates, but they are more robust as they
properly account for the correlations among the parameters.  The reduced
$\chi^2$ of the fit is 1.4. It should be noted that the procedure of
Ref.~\cite{Schmidt84}, which plots the decays on a logarithmic time axis,
fails for this compound spectrum due to the unfortunate position of the
deadtime window which blocks out the peak of the long component in that
formalism. The final results are summarized in \tref{tab:fits} and the
associated fit appears as the solid line in \fref{fig:SFdecays}, with the
dashed and dot-dashed lines representing the individual components.
\begin{table}
\caption{Summary of fit parameters and and associated MINOS errors for the
two measured decay components. See text for discussion.\label{tab:fits}}
\begin{ruledtabular}
\begin{tabular}{lrr}
Parameter & Short & Long \\ \hline
$A$ (decays/$\mu$s)& $45^{+31}_{-16}$ & $1.6^{+0.9}_{-0.7}$ \\
$\lambda$ $(\mu$s$)^{-1}$ & $0.186^{+0.056}_{-0.043}$ &
$0.0161^{+0.0089}_{-0.0055}$ \\ 
$\tau (\mu$s) & $5.4^{+1.6}_{-1.2}$ & $62^{+32}_{-22}$ \\ 
\end{tabular}
\end{ruledtabular}
\end{table}
Relative cross sections between the two states are obtained by integrating
the decay functions from $t=0$ to $t=\infty$ in order to account for
residues lost in flight and in the acquisition deadtime window.
Interpolating the $^{206}$Pb($^{48}$Ca,2n)$^{252}$No cross section from
Ref.~\cite{Oganessian01} to the present beam energy, and normalizing to the
observed $^{252}$No decays from reactions with $^{206}$Pb target impurities
results in \thisno\ production cross sections of $\sigma = 12^{+18}_{-4}$ nb
for the short component and $\sigma = 5^{+3}_{-2}$ nb for the long
component.  These cross sections are in reasonable agreement with the
production of \thisno\ observed in the prior recent
studies\cite{Oganessian01,Belozerov03}. However, the isomeric production
reported here is nearly a factor of two larger than that observed in
Ref.~\cite{Belozerov03}.

\subsubsection{Limits on the \thisno\ $\alpha$ branch}
\label{sec:alphadecay} 

Having unambiguously identified all decays as originating from $^{250}$No, a
search for the previously unobserved alpha decay branch was performed. From
the mass tables of Refs.~\cite{Myers96,Liran00,Muntian03} a $Q$-value of
$Q_\alpha \approx 9.1$ MeV should be expected for $\alpha$-decay of the
\thisno\ nucleus.  The 1.1 s $^{246}$Fm daughter is known to $\alpha$-decay
with $Q_\alpha = 8.4$ MeV. An energy window of 6.5--11.5 MeV, which
encompasses the energy range of all alpha decays of interest in the present
measurement, was implemented for the $\alpha$ particles. No decays with the
expected energy were detected within a recoil-decay correlation window of
500 $\mu$s.  This is not surprising, since the Geiger-Nuttal rule would
imply an $\alpha$ partial half-life of the order of 50 ms for the expected
energy.  Competition with a fission branch decaying within 50 $\mu$s would
imply the need for a data set containing thousands of decays, or about ten
times the present data.

However, during runs 22--37, one decay was observed 19.4 $\mu$s after
implant in pixel 19-18. Although this decay has a reasonable time
correlation, the energy is much lower than expected. The energy read from
the front strip (19) was 7519 keV while the back strip (18) registered 6622
keV.  This seemingly large difference could be due to the decay occurring
near an edge of the back strip, resulting in incomplete charge collection
from that side.  There are no known nobelium isotopes with an $\alpha$-decay
energy this low.  This decay energy is nearly 1.5 MeV less than expected
from the mass models, and the Geiger-Nuttal rule for a 19.4 $\mu$s
$\alpha$-decay would require a $Q_\alpha$ value of about 10.5 MeV---nearly 3
MeV higher than observed. However, this event cannot be easily dismissed
since, as shown in \tref{tab:eventrates}, the probability of this
correlation being random is negligible. The deficit in energy may be due to
partial energy loss attributable to an escaping alpha, incomplete charge
collection if the decay occurred near an edge of the strip, or, less likely,
an unexpected large deviation in the
mass surface for $A=250$.

As a check of the $\alpha$ detection sensitivity in the present experiment,
as well as to ensure that the transport efficiency through the FMA is
understood, a search for longer-lived alpha decays produced from reactions
with the heavier lead isotopes in the target was undertaken. Assuming equal
transport efficiency through the FMA for all nobelium residues, 100\%
efficiency for detecting fission events, 50\% efficiency for $\alpha$
particle detection with full-energy, and cross sections interpolated from
Ref.~\cite{Oganessian01}, an estimate of 5 or 6 total full-energy alpha
decays from $^{251,252}$No should be expected (these are the only isotopes
with lifetimes less than the 10 s time window). Also, within a 10 s
correlation window, the measured event rates should result in $\sim$0.5
random correlations (0 or 1) summed over all pixels in the detector for the
entire experiment.  All such observed recoil-$\alpha$ correlations are
listed in \tref{tab:alphas}.  The anomalous $\alpha$ event described above
is labeled as decay \#0.  From the remaining decays one is assigned as being
random (decay \# 1) since, although the correlation time is $^{252}$No-like,
the measured energy is very close to that of $^{212}$Rn, one of the
long-lived backgrounds resulting from the set-up reaction described in
section~\ref{sec:resultsCaYb}. However, this event could also correspond to
a $^{252}$No decay in which some of the $\alpha$ energy was lost due to an
escape of the particle. The other six decays match expected correlations for
$^{251,251m,252}$No. Opening the time correlation windows further (200 s for
implant-decay, 5000 s for decay-decay), one double-$\alpha$ correlation
chain consistent with the 
$^{248}$Fm$\rightarrow{}^{244}$Cf$\rightarrow{}^{240}$Cm sequence was
observed, assuming that the first $\alpha$ from a $^{252}$No implant was
missed.  Three $\alpha$ decays characteristic of $^{254}$No and one fission
event from $^{252}$No were also recorded, in agreement with expectations.
The number of other correlations in these longer time windows is consistent
with expected random rates and the decay energies/times do not match known
decays, confirming their random character.  

Of the six real decay events with a correlation time between 1 ms and 10 s,
one was correlated to a second (daughter) decay: this is decay 7a \& 7b in
\tref{tab:alphas}. The mass calculated from the focal plane position of the
implant is 250.6.  An implant-decay correlation time of 900 ms for the
parent also favors mass 251 as the progenitor.  The authors of
Ref.~\cite{Hessberger04} showed that $^{251}$No has a low-lying isomer which
decays to an isomeric state in $^{247}$Fm. The $Q_\alpha$ values for the
parent and daughter decays through the isomers are 8805 keV and 8305 keV,
respectively, with lifetimes of 930 ms and 4.3 s. These values match the
observed ones (using the energy from the back strip, which is the largest
signal) quite well.  

\newcommand{\phm}{\phantom{m}}
\begin{table*}
\caption{Observed $\alpha$ decays correlated with a recoil.
For events with a multiplicity $> 1$, the quoted energy is the sum among
the strips. The strip with the largest energy signal is used for pixel
identification  (F = Front side, B = Back side). The last column provides the
isotope assigned to each decay based on the energy and time characteristics in
conjunction with the mass number determined from the focal plane.
\label{tab:alphas}}
\begin{ruledtabular}
\begin{tabular}{rrr|crrrr|rr}
 &&& & \multicolumn{4}{c|}{Energy (keV)} &  \\ \cline{5-8}
\multicolumn{1}{c}{Decay}& \multicolumn{1}{c}{Run} & Event & 
\multicolumn{1}{c}{Strip \#} &
\multicolumn{2}{c}{Sh.\ Amp} & \multicolumn{2}{c|}{Del.\ Amp} & 
\multicolumn{1}{c}{Correlation} &
\multicolumn{1}{c}{Interpretation} \\ 
\multicolumn{1}{c}{\phantom{\#}\#} & \multicolumn{1}{c}{group} & number & 
F-B & F & B & F & B & 
\multicolumn{1}{c}{dt} & \\ 
\hline
0 & 22-37 & 5766611 & 19-18 &  515 &  --  & 7519 & 6622 & 19.4 $\mu$s & unknown\\\hline
1 & 22-37 & 4494828 & 23-24 & 6937 & 6932 & 6940 & 6945 & 2.38 \phm s & random \\
2 & 38-43 &  373050 & 16-18 & 8637 & 8513 & 8629 & 8653 & 114.07 ms & $^{251}$No \\
3 & 38-43 &  713689 & 25-36 & 8700 & 8661 & 8678 & 8638 & 1.04 \phm s & $^{(251,252)}$No\\
4 & 38-43 &  803545 & 16-10 & 8654 & 8591 & 8645 & 8759 & 28.26 ms & $^{(251,252)}$No\\
5 & 38-43 & 1124662 & 18-37 & 8786 & 8681 & 8263 & 8637 & 4.46 \phm s & $^{252}$No\\
6 & 38-43 & 2235165 & 35-26 & 8629 & 8659 & 8620 & 8676 & 2.94 \phm s & $^{252}$No\\
7a &38-43 & 3806836 & 31-05 & 8183 & 8521 & 8502 & 9002 & 904.91 ms & $^{251m}$No \\
7b &38-43 & 3807197 & 31-05 &  --  & 8100 & 6637 & 8489 & 9.60 \phm s & $^{247m}$Fm \\
\end{tabular}
\end{ruledtabular}
\end{table*}

\section{Discussion} \label{sec:discussion}
The ramifications of the present measurement on the structure of \thisno\
are now discussed.  At first, a simple barrier-penetration model can be
used, assuming a barrier with the shape of an inverted harmonic oscillator
and a transmission coefficient (penetrability) given by the Hill-Wheeler
formula
\[ 
  p = (1 + \exp[(2\pi/\hbar\omega)(V-E)])^{-1}.
\]
In addition, the barriers of $^{252}$No and \thisno\ are assumed to have the
same curvature, but different heights.  Within such a framework, the drop in
SF lifetime of 6 orders of magnitude from $^{252}$No to \thisno\ would imply
a lowering of the barrier height of $\sim$1 MeV.  This is remarkable since
many calculations of fission barriers in this
region~\cite{Moller74,Smolanczuk95,Mamdouh01,Koura01} predict a change of at
most 0.6 MeV, usually less. However, there is no \emph{a priori} reason to
expect the barriers of $^{252}$No and \thisno\ to have the same shape.  The
dramatic drop in lifetime could be due to an explicit narrowing of the inner
barrier or to a manifestation of the disappearance of a second minimum in
the potential (effectively making the barrier narrower).  Regardless of the
cause, if this trend would continue to the next even-even nobelium isotope,
$^{248}$No, the SF lifetime would approach the picosecond range and thus be 
technically very difficult to measure.

New information regarding the nuclear structure of \thisno\ can be obtained
from the present data.  Since this nucleus is even in both proton and
neutron number, the ground state will correspond to a $J^\pi$=$0^+$
configuration. Assuming similar structure to $^{252}$No and $^{254}$No,
\thisno\ will have a sizable prolate deformation.  Consequently, the lowest
excited states will most likely correspond to a collective, rotational band
whereas intrinsic excitations will arise from pair-breaking to form
2-quasiparticle configurations. The observed cross sections reveal a ratio
of $\sigma_{\mathrm{short}}/\sigma_{\mathrm{long}} = 2.45$.  That is, 29\%
of the cross section feeds a state associated with the long decay component.
Ghiorso~\cite{Ghiorso73} found not only that the 0.28~s isomer in $^{254}$No
was populated by 20--30\% of the cross section, but also that 30\% of the
$^{250}$Fm production fed an isomeric level in that nucleus as well. The
largest reported ratio of isomer to ground state production in an even-even
nucleus occurs in $^{270}$Ds~\cite{Hofmann01}, where the cross section is
evenly divided between the ground and isomeric states.  Thus, it is
suggested that the more intense short component ($t_{1/2}=3.7$ $\mu$s)
corresponds to the ground state, while the weaker component associated with
the long decay lifetime ($t_{1/2}=43$ $\mu$s) is an isomeric state fed with
a smaller cross section.

The factor of ten difference in lifetime between the two states should be
noted, and its implications for the nature of the isomeric state considered.
The longer SF activity associated with the isomer reported here can result
from direct fission from the excited level, fission from the ground state
that is $\gamma$-delayed by a $K$-forbidden transition from the isomer to
the ground state rotational band, or a mixture of these modes.  If the
longer-lived excited state does indeed feed the ground state there should,
in principle, be a growth and decay aspect to the overall decay spectrum.
Unfortunately, the sensitivity of the present experiment is insufficient to
definitively discern between independent activities and a feeding scenario.
In either case, the SF partial half-life of the isomer must be
$t^{SF_{m}}_{1/2}\ge 43$ $\mu$s.  Both isomer decay possibilities (direct
fission or $K$-forbidden $\gamma$-decay) are discussed below.  The $N=148$
isotone, $^{244}$Cm, has a $K^\pi = 6^+$ isomer at 1040 keV~\cite{Akovali03}
with a half-life of $t^\gamma_{1/2} = 34$ ms.  This nucleus can be used
for guidance in understanding the structure of \thisno.

The systematic review of spontaneous fission by Bj{\"o}rnholm and
Lynn~\cite{Bjornholm80} notes that the presence of an unpaired particle in
the Pu to Fm region hinders ground state fission by as much as $10^5$, and
isomeric fission by $5 \times 10^3$. Furthermore, they note that odd-odd
nuclei are hindered by factors close to that of the product of the two
odd-$A$ neighbors.  This hindrance is attributed to either an effective
increase in barrier height due to reduced pairing, the possible necessity to
conserve the $K$ quantum numbers of all unpaired particles during the decay
process, or a combination thereof. Trends in the rutherfordium isotopes can
provide guidance for estimating these effects in the nobelium isotopes. The
odd Rf nuclei have SF hindrances of the order of $10^3$ relative to their
even neighbors. Considering that the low-lying intrinsic excited states in
\thisno\ are most likely two-quasiparticle excitations, a SF hindrance of at
least $(10^3)^2=10^6$ can be expected.  A more qualitative statement
regarding the 2-quasiparticle SF hindrance factor, $H$, can be obtained from
the observed lifetime ratios. The mean lifetime, $\tau$, is inversely
proportional to the penetrability, $p$, given above. The ratio of the
lifetimes can then be written as
\[
\frac{\tau^{SF}_{m}}{\tau^{SF}_{gs}} = \frac{p_{gs}}{p_{m}}\times H = PH, 
\] 
where $P$ is the ratio of penetrabilities due to excitation and $H$ is the
hindrance factor from other quantum-mechanical considerations.  For
$t^{SF}_{1/2} \ge 43$ $\mu$s, the value of $PH \ge 11$ from the present data
indicates that the hindrance nearly balances the effect of increased
penetrability.  Vandenbosch and Huizenga~\cite{Vandenbosch73} noted that for
estimating the contribution of fission to competing decay modes from excited
states, the fission lifetime is typically expected to decrease by a factor
of approximately $10^8$ per MeV of excitation. Recalling the 1 MeV
excitation known in $^{244}$Cm, a rough estimate of $P=10^{-8}$ can be made
for \thisno, implying a SF hindrance of $H \ge 10^9$. However, using only
700 keV as thermal excitation, $E-E_{\mathrm{yrast}}$, one finds a more
conservative value of $H \ge 10^{6}$. Given the uncertainties in such an
extrapolation, these values are consistent with that obtained from the naive
quasiparticle estimate. It should be noted that for thinner barriers, the
relative increase in penetrability is less than $10^8$ which would also lead
to a somewhat smaller value for $H$. In any event, a large SF hindrance
factor has been observed for this configuration. A strong fission branch
from a $K$-isomer would be unique. The authors of Ref.~\cite{Hall89}
determined a very small spontaneous fission branch ($\sim 10^{-5}$) from the
$K^\pi = 7^-$ isomer in $^{256}$Fm from two observed SF decays.  That work
reported a value of $PH \sim 10^{-7}$, which is vastly different from that
found in the present experiment.  This large difference could be reconciled
if there are unexpected effects in the systematics of fission barriers and
barrier shapes in the superheavy region.

High-$K$ isomers have been reported in
$^{254}$No\cite{Ghiorso73,Tandel06,Herzberg06}, $^{252}$No\cite{Robinson06},
and $^{251}$No\cite{Hessberger04}. The present data provide evidence to
suggest that another $K$ isomer in the No isotopes has been discovered. To
further explore this possibility, the deformed shell model in the nobelium
region has been used to determine what $K$ isomers may be expected. As
mentioned earlier, one other even-even $K$ isomer in the $N=148$ isotones is
known: a 34 ms level in $^{244}$Cm is assigned to a $K^\pi = 6^+$ state at
1040 keV~\cite{Akovali03}.  Table \ref{tab:Kisomers} compares the half-lives
and, where available, production ratios of ground and isomeric states for
all reported $K$ isomers in the even-even transuranic nuclei.
\providecommand{\phd}{\phantom{0}}
\begin{table*}
\caption{Summary of low-lying $K$ isomers identified thus far in the
transuranic nuclei. The half-life for each isomeric ($t^m_{1/2}$) and ground
($t^g_{1/2}$) state as well as the associated ratios are provided in columns
2--4.  Column 5 lists the cross section ratio of isomer to ground state
configurations when both were produced via fusion-evaporation reactions, and
the assigned spin and parity of the isomeric levels are given in column 6.
All isomeric lifetimes listed are for $\gamma$ decay.
\label{tab:Kisomers}}
\begin{ruledtabular}
\begin{tabular}{lrrlccc}
Isotope & 
\multicolumn{1}{c}{$t^{m}_{1/2}$} & 
\multicolumn{1}{c}{$t^{g}_{1/2}$} & 
\multicolumn{1}{c}{$\frac{t^{m}_{1/2}}{t^{g}_{1/2}}$} &
$\frac{\sigma_{m}}{\sigma{g}}$ & $K^\pi$ & Ref. \\ \hline
$^{244}$Cm$_{148}$ & 34\phm\phm ms & 18.1 \phm y & $\phd 6.0\times 10^{-11}$ & & $6^+$ & 
\cite{Akovali03} \\ 
$^{250}$Fm$_{150}$ & 1.8\phm \phm s   & 30\phm min & $\phd 1.0 \times 10^{-3}$ & 0.3 & ($7^-$,$8^-$) &
\cite{Ghiorso73}\\
$^{256}$Fm$_{156}$\footnote{A small SF branch was also reported for this 
isomer decay with a partial half-life of 0.8 ms.} 
& 70\phm\phm ns   & 172\phm min & $\phd 6.8 \times 10^{-12}$ && $7^-$ &
\cite{Hall89}\\ 
$^{250}$No$_{148}$ & 43\phm\phm $\mu$s & 3.7\phm$\mu$s & 11.6 & 0.4 & ($6^+$) &
this data \\ 
$^{254}$No$_{152}$ & 0.28 \phm s & 51 \phm\phm s & $\phd 5.5\times 10^{-3}$ & 0.2--0.4 & $8^-$ &
\cite{Ghiorso73,Tandel06,Herzberg06}\\ 
$^{270}$Ds$_{160}$ & 6.0\phm ms & 1.0\phm ms & \phd 6.0 & 1.0 & $9^-$,$10^-$ &
\cite{Hofmann01}\\ 
\end{tabular}
\end{ruledtabular}
\end{table*}
Predictions of the excitation energy, spin and parity assignments for states
in $^{250}$No have been obtained using multi-quasiparticle blocking
calculations. Specifically, the set of single-particle orbitals originating
from three oscillator shells (N=4, 5 and 6 for protons and N=5, 6 and 7 for
neutrons) were taken from the deformed Woods-Saxon model with the universal
parameters of Ref.~\cite{Cwiok87}, and deformation parameters
$\beta_{2}=0.235$, $\beta_{4}=0.032$ and $\beta_{6}=-0.035$
~\cite{Moller95}.  The pairing correlations were treated using the
Lipkin-Nogami prescription~\cite{Nazarewicz90} with fixed strengths of
G$_{\pi}$=24/A MeV and G$_{\nu}$=17.8/A MeV that, on average, reproduce the
pairing gaps in the region, as reported in Ref.~\cite{Moller95}. The
calculations predict a particularly favored 2-quasineutron state at 1050 keV
(note that the energy does not include a correction for the additional
effect of the spin-spin residual interaction that is of the order of 100
keV) with a configuration $\nu^{2}$(5/2$^+$[622],7/2$^+$[624])$_{6^{+}}$.
This is the same configuration as the $^{244m}$Cm isotone and is a candidate
for the isomer in $^{250}$No reported here.

\begin{figure}
\includegraphics[width=.8\columnwidth]{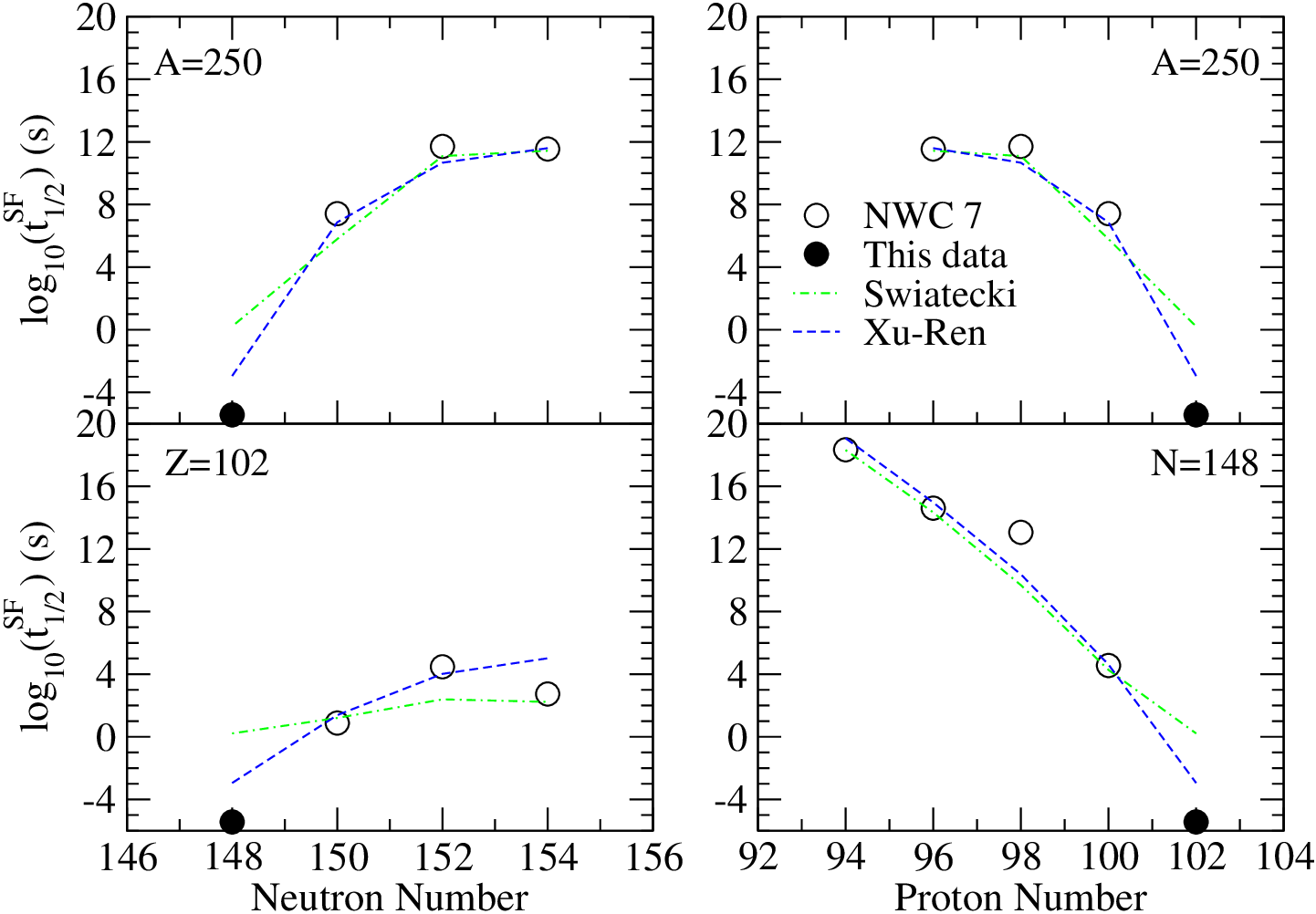}
\caption{Systematics of spontaneous fission (SF) decay lifetimes.  The
logarithm of the SF partial half-life is given as a function of neutron and
proton number for isotopes, isotones, and isobars in the mass 250 region.
See text for discussion.\label{fig:fissionsyst}}
\end{figure}
\begin{figure}
\includegraphics[width=.8\columnwidth]{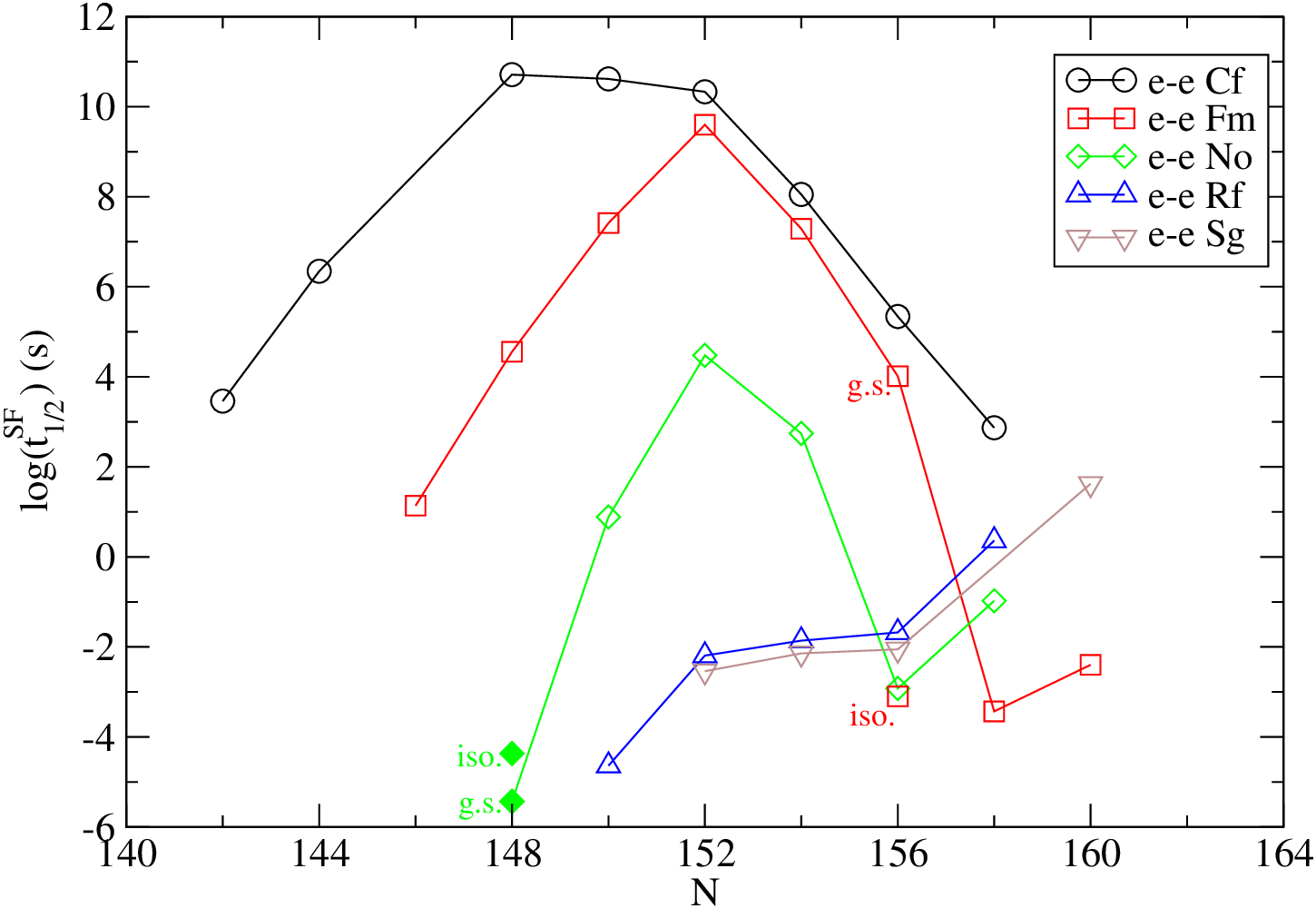}
\caption{Logarithm of partial spontaneous fission lifetimes of even-even
nuclei in the $N=152$ region.\label{fig:transCflifetimes}}
\end{figure}

Finally, the SF systematics of even-even nuclei in the $A=250$ region can be
compared with various semi-empirical treatments. The top panels in
\fref{fig:fissionsyst} are for the $A=250$ isobaric chain. The bottom panels
show the $Z=102$ isotopic and $N=148$ isotonic trends.  Open circles are the
values obtained from Ref.~\cite{Audi03}, excluding systematic
extrapolations.  The filled symbols represent the 3.7 $\mu$s half-life from
the present data. The dot-dashed lines show the predictions of the
six-parameter semi-empirical model of Swiatecki~\cite{Swiatecki55}, which
assumes a smooth trend of the half-lives as a function of $Z^2/A$. The
parameters were obtained from a fit to a limited number of nuclei available
at the time of publication.  The dashed lines follow the modern
parameterization of Xu and Ren~\cite{Xu05}, which is based on the formula of
Ref.~\cite{Swiatecki55}, but contains only four parameters which are fitted
to more data, including 33 even-even isotopes from Th to Cf. The ground
state half-life measured in the present experiment is shorter than any of
the predictions, but clearly favors the newer parameterization of
Ref.~\cite{Xu05}. Figure~ \ref{fig:transCflifetimes} illustrates the
measured partial SF half-lives for the even-even isotopes in the $N=152$
region from Cf to Sg ($Z$ = 98--106). The No nuclei have the sharpest drop
on either side of $N=152$, which appears to indicate that nobelium gains the
most stability from this deformed shell closure.

\section{Summary and Conclusion} \label{sec:summary} 
The $^{204}$Pb($^{48}$Ca,2n) reaction has been measured with an isotopically
enriched target.  Two spontaneous fission activities were observed and
unambiguously linked to the $^{250}$No compound nucleus. The shorter decay
($t_{1/2}=3.7^{+1.1}_{-0.8}$ $\mu$s) is assigned to the ground state of
\thisno. The longer decay ($t_{1/2}=43^{+22}_{-15}$ $\mu$s) is attributed to
a $K^\pi = 6^+$ isomeric state of \thisno\ based on: 1) a smaller production
cross section and 2) the prediction of a 2-quasineutron state around 1050
keV based on multi-quasiparticle blocking calculations.  These results
require an update to the latest compilations~\cite{Tuli05}, which assign the
longer lifetime to \lightno\ rather than to $^{250m}$No. Including the
measurements of Ref.~\cite{Belozerov03} with the present data yields global
average half-lives of $t_{1/2}=4.2^{+1.2}_{-0.9}$ and 
$t_{1/2}=46^{+22}_{-14}$ $\mu$s for these two states in \thisno. There is no
evidence for the 250 $\mu$s SF lifetime originally reported in 1975
~\cite{TerAkopyan75}. The present data cannot distinguish whether the isomer
decays via spontaneuous fission directly or proceeds through a $K$-forbidden
$\gamma$-decay to the ground state, which proceeds to fission. In either
case, a very large SF hindrance factor has been observed for this
two-quasiparticle configuration.  The presence of an isomeric level that is
longer-lived than the ground state is also notable.

The search for an $\alpha$-decay branch was inconclusive.  A single observed
$\alpha$-like decay, with a correlation time of 19.4 $\mu$s, places an upper
limit on the $\alpha$ branch of 1.8\% for the ground state and 4.4\% for the
isomeric state. This would imply partial lifetimes, $t_{1/2}^\alpha$, of 205
$\mu$s and 975 $\mu$s for the ground and isomeric states respectively.
(Inclusion of the observations of Ref.~\cite{Belozerov03} sets limits of 1.2\%
and 2.9\% for the two decays.) The anomolously low energy observed for the
single $\alpha$-like decay bears further investigation.  

The dramatic decrease in half-life from $^{252}$No to $^{250}$No suggests
that the No fission barriers decrease rapidly with neutron number to the
extent that in the next even-even nobelium isotope, $^{248}$No, the fission
barrier may disappear or be negligible.  However, since no odd-$A$ nobelium
nucleus has been observed to decay via fission, the hindrance factors must
be very large and it may be possible to produce and observe the decay of
$^{249}$No in a future experiment. Observation of the $\alpha$ decay of
$^{249}$No would also provide further input into the trends in the mass
surface in this region.  If \lightno\ were to also have an isomeric level
that could be observed via $\alpha$-$\gamma$ or $\alpha$-$e^-$ correlations,
the possibility may exist to infer the single-particle configuration of the
\thisno\ isomer with greater certainty.

\section{Acknowledgments} 
The authors would like to thank K.-H. Schmidt for helpful discussions
regarding lifetime extraction, J.~P. Greene for making the targets, and the
ATLAS operations staff for providing stable beams and equipment during the
experiment.  This work was funded by the U.S. Department of Energy, Office
of Nuclear Physics, under contract nos.\ W31-109-ENG-38 (ANL),
DE-FG02-91ER-40609 (Yale), and DE-FG02-94ER40848 (U.\ of Mass., Lowell).


\end{document}